\documentclass[]{interact}

\usepackage{epstopdf}
\usepackage[caption=false]{subfig}

\usepackage[numbers,sort&compress]{natbib}
\bibpunct[, ]{[}{]}{,}{n}{,}{,}

\theoremstyle{plain}

\theoremstyle{definition}

\theoremstyle{remark}

\begin{document}


\title{Kinetic modelling of economic markets with individual and collective transactions}

\author{
\name{Chuandong Lin\textsuperscript{a} and Lijie Cui\textsuperscript{b} \thanks{CONTACT Lijie Cui. Email: cuilj7@mail.sysu.edu.cn}}
\affil{\textsuperscript{a}Sino-French Institute of Nuclear Engineering and Technology, Sun Yat-sen University, Zhuhai, 519082, China; \textsuperscript{b}Socioeconomic Survey Team of Guangming District, Shenzhen, 518107, China}
}

\maketitle

\begin{abstract}
Two kinetic exchange models are proposed to explore the dynamics of closed economic markets characterized by random exchanges, saving propensities, and collective transactions. Model I simulates a system where individual transactions occur among agents with saving tendencies, along with collective transactions between groups. Model II restricts individual transactions to agents within the same group, but allows for collective transactions between groups. A three-step trading process--comprising intergroup transactions, intragroup redistribution, and individual exchanges--is developed to capture the dual-layered market dynamics. The saving propensity is incorporated using the Chakraborti--Chakrabarti model, applied to both individual and collective transactions. Results reveal that collective transactions increase wealth inequality by concentrating wealth within groups, as indicated by higher Gini coefficients and Kolkata indices. In contrast, individual transactions across groups mitigate inequality through more uniform wealth redistribution. The interplay between saving propensities and collective transactions governs deviation degree and entropy, which display inverse trends. Higher saving propensities lead to deviations from the Boltzmann--Gibbs equilibrium, whereas specific thresholds result in collective transaction dominance, producing notable peaks or troughs in these metrics. These findings underscore the critical influence of dual-layered market interactions on wealth distribution and economic dynamics.
\end{abstract}

\begin{keywords}
Kinetic model; multi-agents; economic transaction; wealth distribution
\end{keywords}

\section{Introduction}\label{SecI}

The economic market is a fundamental aspect of human society, drawing significant attention from economists, mathematicians, and physicists around the world \cite{Joohyun2017JS,Mantegna1999}. Understanding economic dynamics is crucial, as it directly influences society development and is closely related to the severe social challenges arising from wealth inequality \cite{Yakovenko2009RMP}. Given the practical significance, economic transitions, as complex social phenomena, have been extensively examined through a variety of theoretical and empirical approaches. During the past century, significant progress has been made in the development of economic theories. As early as the 1890s, Pareto introduced the power law function to describe the distribution of wealth in society \cite{Pareto1897Book}. In 1912, the Gini coefficient, in relation to the Lorenz function, was introduced as a measure of inequality in wealth distribution \cite{Gini1912JEI}. Later, the Kuznets curve was proposed under the hypothesis that market forces first increase and then decrease economic inequality during the development of an economy \cite{Kuznets1955AER}. Notably, in 2001, Dr\u{a}gulescu and Yakovenko demonstrated that the steady money distribution takes the Boltzmann--Gibbs exponential form in an ideal free economic system \cite{Dragulescu2001EPJB}. In 2014, as a special point in the Lorenz curve, the Kolkata index was proposed to depict the wealth inequality state that a large portion of wealth is held by a small fraction of the population \cite{Ghosh2014PA}.

Moreover, in the past few decades, empirical investigations of economic transactions have also been widely carried out by analyzing economic data through statistical analysis and numerical simulation. Note that numerical research, or computational modelling, has facilitated academic research on a variety of economic phenomena. Specifically, some intriguing kinetic exchange models for wealth transactions between traders have been proposed \cite{Chakraborti2000EPJB,Chakrabarti2009PA,Cha2011CPC,Fernandes2020EPJB,Munoz2022Chaos}. In 2000, Chakraborti and Chakrabarti introduced the kinetic model of CC to study economic transactions among individual agents driven by self-interest \cite{Chakraborti2000EPJB}. In 2004, Chatterjee et al. constructed a model of trading markets where an inhomogeneous saving factor is distributed within the population \cite{Chatterjee2004PA}. In 2013, Cerd\'{a} presented an econophysics model based on the lattice Boltzmann automaton, and simulated income distribution with respect to tax regulation \cite{Cerda2013MCM}. In 2014, Heinsalu and Patriarca introduced the HP model, which incorporates modified basic dynamics to represent instantaneous wealth exchanges and employs a general probabilistic criterion to govern the trading decisions of economic agents. Interestingly, the HP model's framework of immediate exchange establishes connections to the kinetic models developed by Dr\u{a}gulescu and Yakovenko \cite{Dragulescu2001EPJB}, Chakraborti and Chakrabarti \cite{Chakraborti2000EPJB}, and Angle \cite{Angle1986SF}. In 2017, Boghosian et al. introduced a bias in favor of the wealthier agent of wealth-attained advantage model, leading to the phenomenon of ``wealth condensation" when the bias exceeds a certain critical value \cite{Boghosian2017PA}. In 2018, Vermeulen utilized Monte Carlo evidence and found that adding only a few extreme observations to wealth surveys is sufficient to remove the downward bias \cite{Vermeulen2018RIW}. In the same year, D$\ddot{u}$ring et al. introduced and investigated optimal control strategies for kinetic models for a multi-agent economic system \cite{During2018TEPJB}. In 2019, Li et al. introduced a stochastic agent-based model for binary transactions in economic systems, allowing for the possibility of agents holding negative wealth \cite{Li2019PA}. The following year, Cui and Lin developed a lattice gas automaton framework to model individual transactions, incorporating factors such as income tax, Matthew effect, and charity \cite{Cui2020Entropy}. In 2021, they further proposed a streamlined one-dimensional lattice gas automaton designed to study random trade between individual agents, accounting for scenarios both with and without saving propensities \cite{Cui2021PA}. In the same year, Sargent et al. measured Gini coefficients, fractile inequalities, and tail power laws, concluding that wealth is less evenly distributed across people than labor earnings \cite{Sargent2021PNAS}. In 2022, Goswami investigated the dynamics of agents below a threshold line in modified kinetic wealth exchange models \cite{Goswami2022PTRSA}.

Computational models have proven highly effective in analyzing economic markets by incorporating a wide range of personal and societal factors \cite{Patriarca2004PA,Patriarca2007EPJB,Chakraborti2008PJOP,Patriarca2010EPJB,Chakraborti2011QF,Heinsalu2014EPJB,Chatterjee2017PA,Patriarca2017book1,Patriarca2017book2,Hu2006EPJB,Ghosh2016PA,Kulp2019Chaos,Cui2023IJMPC,Cui2023EPL}. In 2004, Patriarca et al. demonstrated that wealth distribution follows a Boltzmann--Gibbs pattern in an economic market where saving propensity is absent, transitioning to a gamma distribution in systems with nonzero saving interests \cite{Patriarca2004PA}. In 2007, Patriarca et al. utilized a statistical multi-agent model to explore wealth exchange dynamics and investigated how the relaxation process is influenced by varying saving propensities \cite{Patriarca2007EPJB}. In the same year, Hu et al. found that the relationship between personal wealth and its connectivity is a possible mechanism for the emergence of the Matthew effect in the economy \cite{Hu2006EPJB}. In 2016, Ghosh et al. discovered that the equilibrium wealth distribution could exhibit either a unimodal or bimodal pattern, determined by the interplay between the saving propensities and the relative population ratios of the two agent groups \cite{Ghosh2016PA}. In 2019, Kulp et al. employed the tax and redistribution models to investigate their influence on the wealth distribution by calculating the Gini coefficient \cite{Kulp2019Chaos}. In 2023, Cui and Lin employed a lattice gas automaton to examine how the saving behaviors within two distinct agent groups and their relative population sizes influenced the resulting wealth distributions \cite{Cui2023IJMPC}. That same year, Cui et al. constructed a kinetic model, revealing that both economic profit and Matthew effect play significant roles in broadening wealth distribution and intensifying wealth inequality \cite{Cui2023EPL}.

Previous numerical studies have primarily concentrated on modelling economic markets with individual transactions. However, no kinetic model has been developed to simulate economic dynamics in a system with collective transactions. In human society, wealth exchanges are not limited to individual transactions; collective transactions, such as those between companies, countries, or other organizations, also play a significant role in shaping economic systems. The inclusion of collective transactions in kinetic modelling is therefore crucial for capturing the full complexity of economic interactions. Such transactions often involve larger-scale dynamics and can significantly influence the overall distribution of wealth within a system. Recognizing this gap, we aim to develop a kinetic model that integrates both individual and collective transactions. This model enables the investigation of wealth distribution under the combined influence of these two types of transactions, providing a more comprehensive understanding of economic dynamics. By analyzing the interplay between individual and collective exchanges, we can uncover new insights into the mechanisms driving wealth accumulation and inequality in complex economic systems.

\section{Research methodology}\label{SecII}

Let us begin by outlining the research methodology adopted in this study. In the first section, we introduce three kinetic exchange models designed to simulate random transactions within an economic market, incorporating the effects of saving propensities. Following this, we describe the analytical framework used to investigate the characteristics of wealth distribution. Specifically, we employ four key metrics: the Gini coefficient, Kolkata index, deviation degree, and entropy, which collectively provide a comprehensive evaluation of the system's equilibrium states and inequality dynamics.

\subsection{Kinetic model}

In this part, we introduce three kinetic models of wealth exchange. The first model, developed by Chakraborti and Chakrabarti \cite{Chakraborti2000EPJB}, focuses exclusively on individual transactions. The other two models, proposed in this study, extend the framework to incorporate both individual and collective transactions, providing a more comprehensive approach to modelling wealth dynamics in economic systems.

\subsubsection{CC Model}
Consider a closed economic system consisting of $N$ independent agents, each representing an individual. An agent $i$ is characterized by their monetary holdings, denoted as $m_i$, where $i = 1$, $2$, $\dots$, $N$. The agents exchange money with one another through a stochastic process that simulates economic transactions. Each agent is assigned a saving propensity $\lambda$ which represents the fraction of their wealth they choose to retain after a transaction. The saving propensity satisfies $0 \le \lambda \le 1$, where $\lambda=0$ implies no savings (agents fully engage their wealth in transactions), and $\lambda=1$ implies complete saving (no transactions occur). When two agents, $i$ and $j$, are randomly selected for a trading interaction, their monetary holdings are updated according to the following rule:
\begin{equation}
	\left\{
	\begin{array}{l}
		{{m}'_{i}} = {{m}_{i}} - \Delta m \tt{,}  \\
		{{m}'_{j}} = {{m}_{j}} + \Delta m \tt{,}
	\end{array}
	\right.
	\label{ExchangeM}
\end{equation}
where $\Delta m$ represents the transferred amount during the transaction, ${{m}_{i}}$ and ${{m}_{j}}$ denote the monetary holdings of agents $i$ and $j$ before transaction, ${{m}'_{i}}$ and ${{m}'_{j}}$ are the money after the transaction. This model assumes that no debt is allowed, hence all monetary holdings remain non-negative throughout the process. In the CC model \cite{Chakraborti2000EPJB}, the trading volume $\Delta m$ is expressed by
\begin{equation}
	\Delta m=(1-\lambda )[{{m}_{i}}-\varepsilon ({{m}_{i}}+{{m}_{j}})]
	\label{DeltaM_CC}
	\tt{,}
\end{equation}
where $\varepsilon$ is a uniformly distributed random number in the interval $0 \le \varepsilon \le 1$, determining the fraction of the transferable wealth allocated to agent $i$.

\subsubsection{Model I}
We now introduce a kinetic model incorporating both individual and collective transactions, building upon the foundational CC model \cite{Chakraborti2000EPJB}. In this framework, $N$ agents are grouped into $N_g$ distinct groups, with each group consisting of $N/{N_g}$ agents. These groups may represent entities such as companies, countries, or other organizations. The model captures two levels of transactions: individual transactions occurring among agents and collective transactions occurring between groups.

To simulate this dual-layered market dynamic, we divide the trading process into three distinct steps, detailed as follows:

Step 1: Collective transactions among groups

In the first step, we model transactions between groups. The total monetary holdings of each group are calculated by summing the wealth of all agents within each group. Let the total money in groups $p$ and $q$ before the transaction be $M_{p}$ and $M_{q}$, respectively. Suppose the trading volume is $\Delta M$, and the monetary holdings of the groups after the transaction are ${{M}'_{p}}$ and ${{M}'_{q}}$, respectively. The evolution of the monetary holdings is governed by:
\begin{equation}
	\left\{
	\begin{array}{l}
		{{M}'_{p}} = {{M}_{p}} - \Delta M \tt{,}  \\
		{{M}'_{q}} = {{M}_{q}} + \Delta M \tt{,}
	\end{array}
	\right.
	\label{ExchangeGroup}
\end{equation}
where the trading volume $\Delta $ is given by:
\begin{equation}
	\Delta M=(1-\lambda )[{{M}_{p}}-\varepsilon ({{M}_{p}}+{{M}_{q}})]
	\label{DeltaGroup_CC}
	\tt{.}
\end{equation}
Here $\lambda$ represents the saving propensity of the groups, controlling the fraction of wealth retained during the transaction. $\varepsilon$ is a uniformly distributed random variable in the range [$0$, $1$], determining the stochastic distribution of wealth exchange. Mathematically, Eqs. (\ref{ExchangeGroup}) and (\ref{DeltaGroup_CC}) are derived by adapting the individual transaction rules (Eqs. (\ref{ExchangeM}) and (\ref{DeltaM_CC})) to the group level, replacing individual monetary holdings ${{m}_{i}}$ and ${{m}_{j}}$ with group monetary holdings ${{M}_{p}}$ and ${{M}_{q}}$, respectively.

Step 2: Redistribution within groups

After the collective transaction, the monetary holdings of individual agents within each group are adjusted proportionally. Let the monetary holding of agent $i$ in group $p$ before and after redistribution be ${{m}_{i}}$ and ${{m}_{i}}'$, respectively. The redistribution is governed by the following rule:
\begin{equation}
	{{m}_{i}}'=\frac{{{m}_{i}} {{M}'_{p}}}{{{M}_{p}}}
	\tt{.}
\end{equation}
This ensures that the total money in group $p$ after redistribution matches the updated group wealth ${{M}'_{p}}$. The redistribution step preserves the proportional wealth distribution within the group while reflecting the impact of the collective transaction.

Step 3: Individual transactions among agents

The final step involves individual transactions among all agents, governed by the original CC model \cite{Chakraborti2000EPJB}. For a randomly selected pair of agents $i$ and $j$, the transaction rules follow Eqs. (\ref{ExchangeM}) and (\ref{DeltaM_CC}).

\begin{figure}[htbp]
	\centering
	\includegraphics[width=0.7\textwidth]{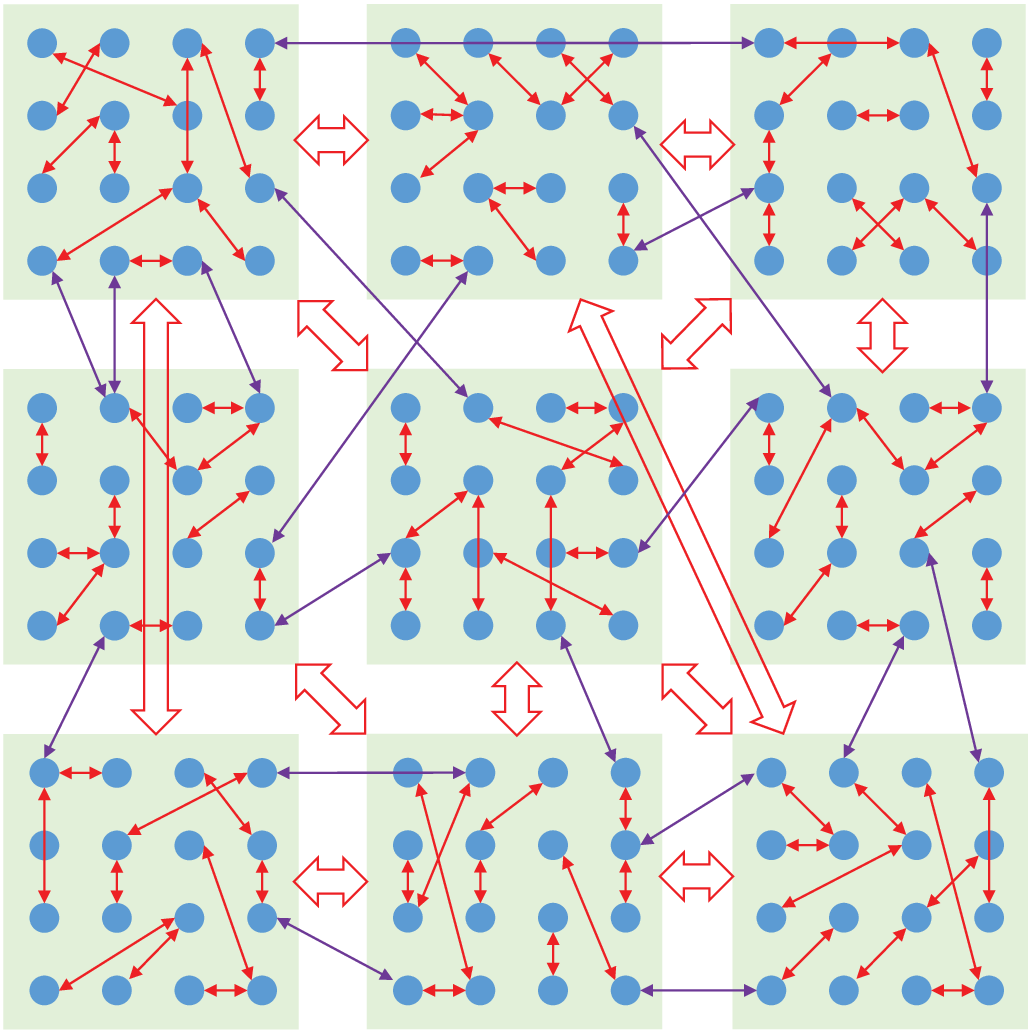}
	\caption{Sketch of Model I: Individual transactions occur among all agents, while collective transactions take place between different groups. Bidirectional arrows represent individual transactions, and rectangular boxes with double-headed arrows represent collective transactions.}
	\label{Fig01}
\end{figure}

Figure \ref{Fig01} provides a schematic representation of Model I. The red bidirectional arrows depict transactions among agents within the same group, while the blue bidirectional arrows represent transactions among agents across different groups. Additionally, the rectangular boxes with double-headed arrows on both ends illustrate transactions occurring between different groups. By integrating both collective and individual transactions, this model effectively captures the dynamic interplay between hierarchical organizational structures and individual behavior in economic systems. The three-step process offers a comprehensive framework for exploring wealth dynamics, inequality, and emergent patterns at both micro (individual) and macro (group) levels.

\subsubsection{Model II}
\begin{figure}[htbp]
	\centering
	\includegraphics[width=0.7\textwidth]{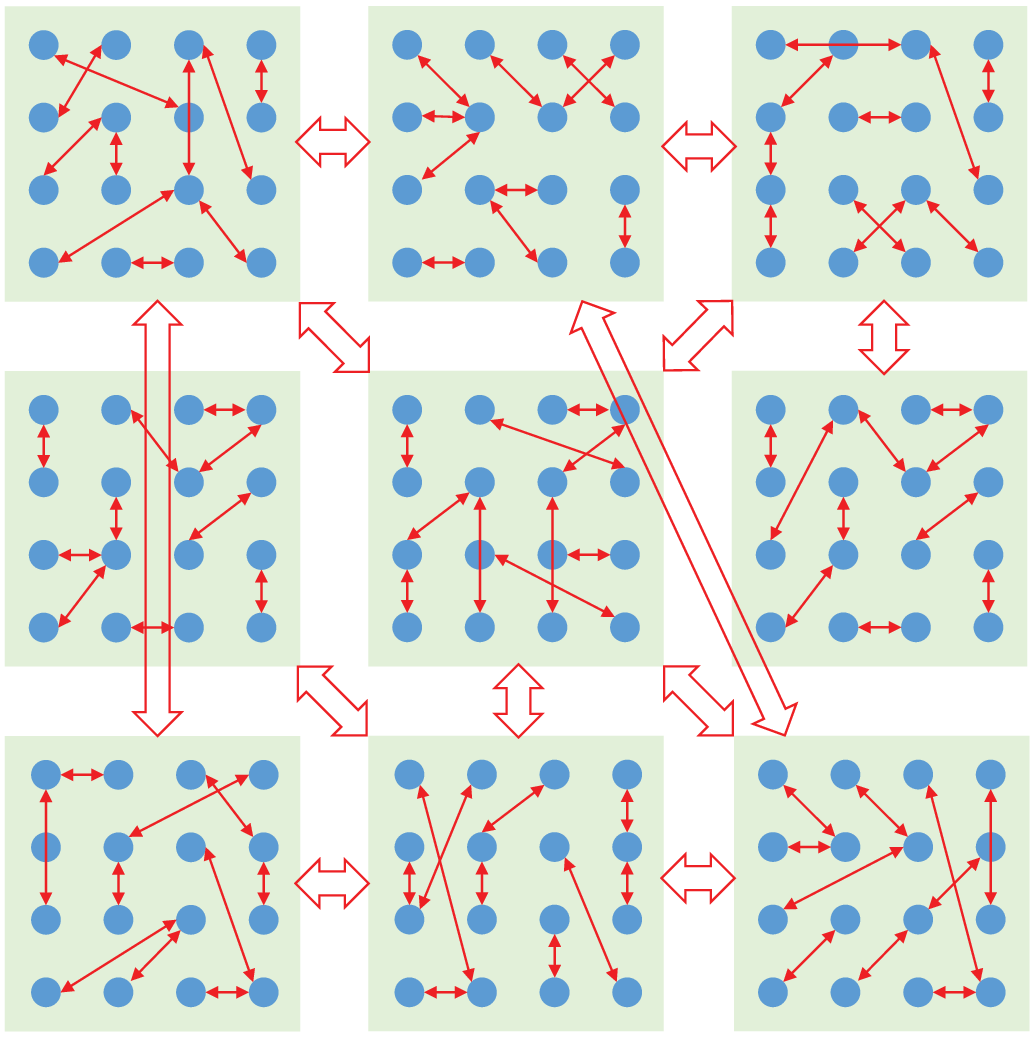}
	\caption{{Sketch of Model II: Individual transactions occur among agents within each group, while collective transactions take place between different groups. Bidirectional arrows represent individual transactions, and rectangular boxes with double-headed arrows represent collective transactions.}}
	\label{Fig02}
\end{figure}

The final model we propose addresses the following scenario: individual transactions occur exclusively among agents within each group, while collective transactions take place between different groups. A schematic representation of this model is shown in Fig. \ref{Fig02}. In comparison to Model I (illustrated in Fig. \ref{Fig01}), this model excludes individual transactions among agents belonging to different groups. Instead, all individual transactions are restricted to agents within the same group.

Model II follows a three-step process similar to Model I. The first two steps remain unchanged: collective transactions among groups and redistribution within groups. However, the third step differs significantly. In Model II, individual transactions occur only among agents within the same group. There are no direct individual transactions between agents from different groups. This distinction highlights the localized nature of interactions in this model.

By limiting individual transactions to intra-group exchanges, Model II emphasizes the impact of localized interactions on wealth dynamics. This approach provides insights into systems where intergroup transactions occur only at the collective level, such as trade between organizations or nations. The model serves as a valuable framework for analyzing wealth distribution, inequality, and emergent patterns under constrained interaction scenarios.

It is important to note that both Models I and II are applicable to human societies but under different historical and social conditions. In earlier times, economic interactions between nations were predominantly characterized by collective transactions, such as trade agreements or exchanges of goods at the national or organizational level. Individual transactions between agents from different countries were rare, primarily due to limited communication and transportation infrastructure, as well as the absence of global markets. In contrast, the modern era of globalization has brought significant changes to the dynamics of economic interactions. Today, both individual and collective transactions occur on a global scale. Advances in technology, particularly the internet, have enabled individuals from different countries to interact and exchange directly. For instance, two people from different parts of the world can easily engage in financial transactions or trade goods and services through online platforms. This shift reflects the profound impact of globalization on economic systems, where barriers to individual interactions are reduced, and the interconnectedness of markets is greatly enhanced. Models I and II, therefore, capture these differing dynamics, providing a framework to understand wealth exchange and interaction patterns in both localized and globalized contexts.

\subsection{Measure quantities}

To quantitatively investigate the impact of various personal and/or social factors (including saving propensity and collective transactions) upon the wealth distribution, we introduce four measure quantities (i.e., the Gini index, Kolkata index, deviation degree, and entropy).

\subsubsection{Gini coefficient} 

To measure wealth inequality in the evolution of a social system, the Gini coefficient can be employed to estimate how far the wealth distribution deviates from an equal state \cite{Gini1912JEI}. Mathematically, the Gini coefficient is defined as
\begin{equation}
	{\rm{G}}=\frac{1}{2{{m}_{0}}}\int_{0}^{\infty }{\int_{0}^{\infty }{f\left( x \right)f\left( y \right)}\left| x-y \right|dx}dy
	\tt{,}
	\label{Gini_Integral}
\end{equation}
where $f = f(m)$ denotes the wealth distribution of individual agents, and ${{m}_{0}}$ the average money per agent. Mathematically, the relationship between $f(m)$ and ${{m}_{0}}$ reads ${{m}_{0}}=\int_{0}^{\infty }{mf\left( m \right)}dm$.

In theory, Eq. (\ref{Gini_Integral}) is suitable for an economic system with an infinite number of agents. For a market with a limited number of agents $N$, the Gini coefficient is expressed by
\begin{equation}
	{\rm{G}} = \frac{1}{2N^{2}{{m}_{0}}}\sum\nolimits_{i=1}^{{N}}{\sum\nolimits_{j=1}^{{N}}{\left| {{m}_{i}}-{{m}_{j}} \right|}}
	\tt{.}
	\label{Gini_Sum}
\end{equation}
In Eq. (\ref{Gini_Integral}) or (\ref{Gini_Sum}), the Gini index is within the range $0 \le {\rm{G}} \le 1$. The Gini index ${\rm{G}} = 0$ means perfect equality, namely, each agent possesses the same wealth. Besides, the Gini coefficient ${\rm{G}} = 1$ expresses maximal wealth inequality among agents in the social system. As the Gini index increases from zero to one, the problem of wealth inequality is exacerbated.

\begin{figure}[htbp]
	\centering
	\includegraphics[width=0.5\textwidth]{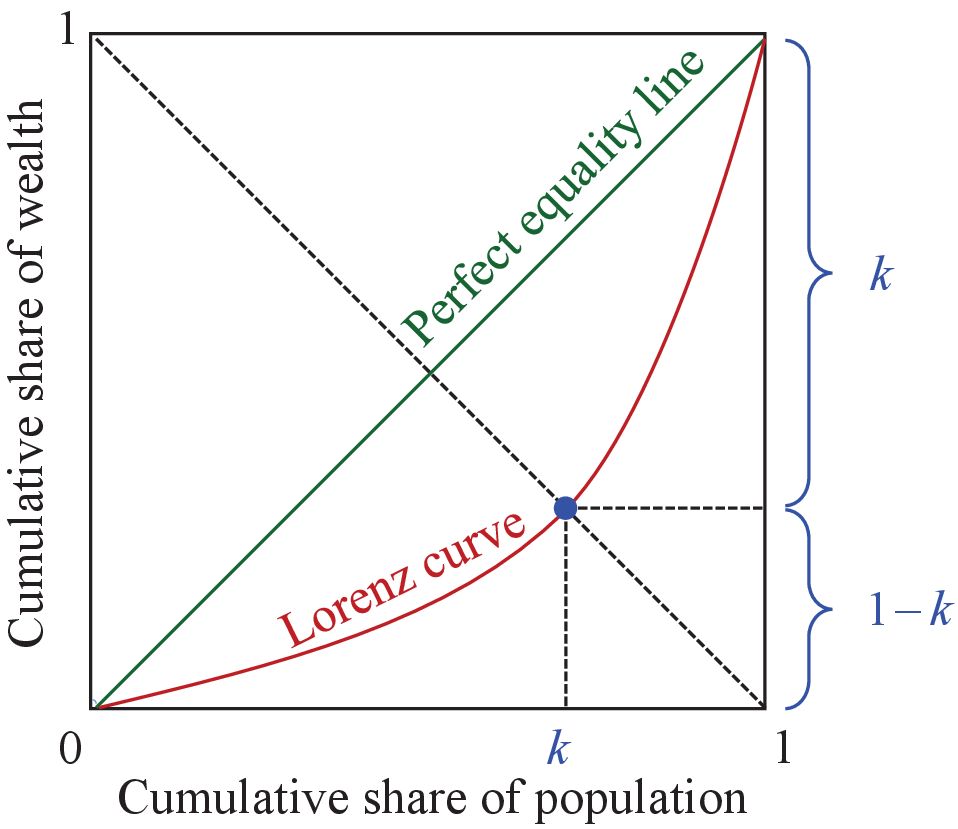}
	\caption{{Lorenz curve and perfect equality line.}}
	\label{Fig03}
\end{figure}

In addition, the Gini coefficient is associated with the Lorenz function curve, which depicts the proportion of overall wealth ($y$-axis) taken by the bottom $(100 x)\%$ of the population (see Fig. \ref{Fig03}). The Lorenz curve is a graph within a unit square, with endpoints at ($0$, $0$) and ($1$, $1$), between which is the line of perfect equality at $45$ degrees as well. Theoretically, the closer a Lorenz curve lies to the equality line, the more equitable the wealth distribution. Mathematically, the Gini coefficient equals the ratio of the area between the Lorenz curve and the equality line to the total area beneath the equality line. Alternatively, the Gini index is twice the area between the Lorenz curve and the equality line.

\subsubsection{Kolkata index}

As another important measure of inequality, the Kolkata index ($k$) also corresponds to the Lorenz curve, having several intriguing and useful properties \cite{Ghosh2014PA,Banerjee2020FIP,Banerjee2023IJMPC}. As shown in Fig. \ref{Fig03}, the Kolkata index is defined as the intersection of the Lorenz curve and the line across the points ($0$, $1$) and ($1$, $0$) with $135$ degrees. In other words, the Kolkata index is a non-trivial point on the Lorenz curve. It means that the $1-k$ proportion of population possess top $k$ fraction of wealth in the society, while the other $k$ people own only the rest $1-k$ wealth.

It should be emphasized that, according to the definition of Kolkata index, the value is in the range of $1/2 \le k \le 1$. An alternative way of measuring the $k$ index is to consider the quantity $K = 2k - 1$, which is within $0 \le K \le 1$ \cite{Chatterjee2017PA}. Mathematically, $K$ index is a normalized $k$ index, and there is a linear relation between them. In fact, the values of $k = 1/2$ and $K = 0$ correspond to the perfect equality. In contrast, $k = 1/2$ and $K = 0$ reflect maximal inequality. With the increasing of $k$ and $K$, the gap between the rich and the poor rises.

\subsubsection{Deviation degree}

\begin{figure}[htbp]
	\centering
	\includegraphics[width=0.4\textwidth]{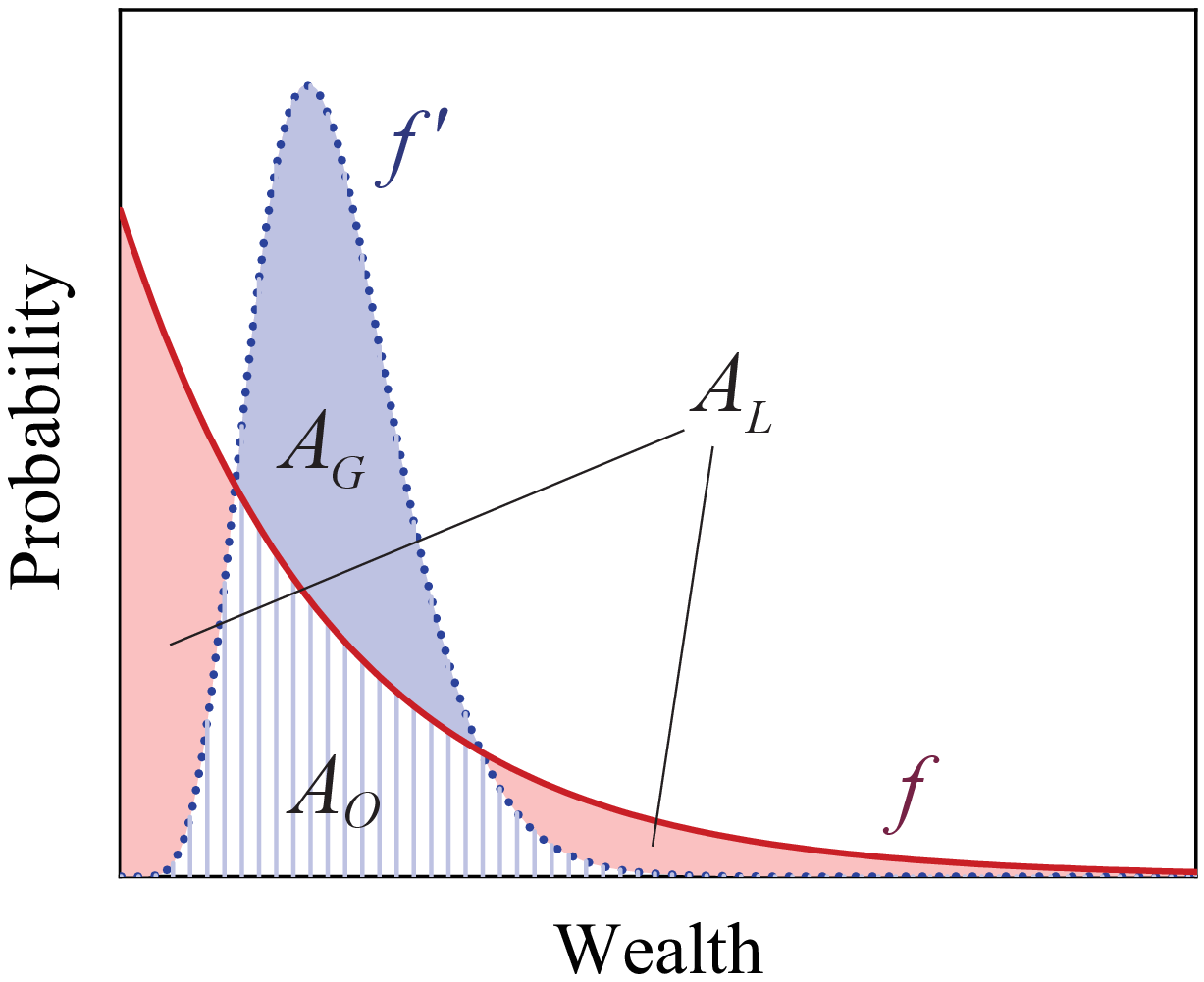}
	\caption{Sketch of two wealth distributions. The solid and dotted lines indicate $f$ and $f'$, respectively.}
	\label{Fig04}
\end{figure}

To quantify the dissimilarity between two arbitrary wealth distributions, we introduce the concept of the deviation degree. This measure provides an intuitive understanding of how different two distributions are. Figure \ref{Fig04} illustrates two example distributions, $f$ and $f'$. Mathematically, the following normalization conditions hold:
\begin{equation}
	A=\int_{0}^{\infty }{f}dm=1
	\tt{,}
	\label{S1}
\end{equation}
\begin{equation}
	{A'}=\int_{0}^{\infty }{{{f}'}}dm=1
	\tt{,}
	\label{S2}
\end{equation}
where $A$ and $A'$ represent the areas under the solid and dotted curves, respectively. These conditions reflect the fact that the total probability of wealth in the system equals one.

The overlap area between the two distributions is denoted as ${A}_{O}$, while the regions where ${f'}<{f}$ and ${f'}>{f}$ are defined as ${A}_{L}$ and ${A}_{G}$, respectively. These areas can be expressed as follows:
\begin{equation}
	{{A}_{L}}={{\left. \int_{0}^{\infty }{\left( {f}-{f'} \right)}dm \right|}_{{f'}<{f}}}
	\tt{,}
	\label{S_L}
\end{equation}
\begin{equation}
	{{A}_{G}}={{\left. \int_{0}^{\infty }{\left( {f'}-{f} \right)}dm \right|}_{{f'}>{f}}}
	\tt{.}
	\label{S_G}
\end{equation}
From the above definitions, the relationships ${{A}_{L}}+{{A}_{O}}={{A}_{G}}+{{A}_{O}}=1$ hold. This leads to the equality:
\begin{equation}
	{{A}_{L}}={{A}_{G}}=\frac{1}{2}\int_{0}^{\infty }{\left| {f'}-{f} \right|}dm
	\tt{.}
	\label{S_L_G}
\end{equation}
The deviation degree $\Delta$ is geometrically defined as the ratio of the non-overlapping area $({{A}_{L}}+{{A}_{G}})$ to the total area $({A}+{{A}'})$:
\begin{equation}
	\Delta =\frac{{{A}_{L}}+{{A}_{G}}}{{A}+{{A}'}}
	\tt{.}
	\label{Delta_definition}
\end{equation}
By combining Eqs. (\ref{S1})-(\ref{Delta_definition}), we find:
\begin{equation}
	\Delta =\frac{\int_{0}^{\infty }{\left| {f'}-{f} \right|}dm}{\int_{0}^{\infty }{\left( {f'}+{f} \right)}dm}=\frac{1}{2}\int_{0}^{\infty }{\left| {f'}-{f} \right|}dm = {{A}_{L}}= {{A}_{G}}
	\tt{.}
	\label{Delta1}
\end{equation}
The deviation degree $\Delta$ quantifies the extent of dissimilarity between two wealth distributions, ranging from $0 \le \Delta \le 1$. When $\Delta =0$, the two distributions are identical. Conversely, $\Delta =1$ indicates complete dissimilarity. As $\Delta$ increases from $0$ to $1$, the two distributions become increasingly different.

This measure has two significant applications \cite{Cui2023EPL}. First, it allows us to study how various factors influence wealth distribution through the control variable method. For example, as an influencing factor changes while other conditions remain constant, the deviation between the distributions increases. Second, $\Delta$ can be used to assess how far a given wealth distribution deviates from an ideal equilibrium distribution in a perfectly free market without external influences \cite{Cui2021PA}.
In this context, $f'$ represents the wealth distribution under real-world economic and social conditions, while $f$ denotes the ideal Boltzmann--Gibbs exponential distribution:
\begin{equation}
	{f} =\frac{1}{{{m}_{0}}}\exp \left( -\frac{m}{{{m}_{0}}} \right)
	\label{Probability1}
	\tt{,}
\end{equation}
which describes the steady-state distribution in a completely free economic system \cite{Dragulescu2001EPJB}. Furthermore, the degree of deviation $\Delta$ can help explain the variation of entropy in economic systems (see Figs. \ref{Fig04} and \ref{Fig05}).

\subsubsection{Entropy}

In order to describe the disorder and random state of the whole economic system, we introduce the entropy that takes the form
\begin{equation}
	S=-\left\langle f\ln f \right\rangle =-\int_{0}^{\infty }{f\ln fdm}
	\tt{,}
	\label{Entropy_Integral_f}
\end{equation}
where the integral is over the whole range of wealth. It is noteworthy that the wealth distribution $f$ denotes the probability density of wealth in the market, as aforementioned. And the entropy of a free closed economic system (without any personal or social factor) reaches the maximum when the monetary distribution takes the Boltzmann--Gibbs form. This maximum is greater than the entropy of an equilibrium or nonequilibrium economic system where the saving propensity, collective transactions or other impact factors exist.

For an economic market comprising a finite number of agents, the formula of entropy can be expressed in the following discrete form
\begin{equation}
	S=-D\sum\limits_{{{f}_{i}}>0}{{{f}_{i}}\ln {{f}_{i}}}
	\tt{,}
	\label{Entropy_Sum_fi}
\end{equation}
where $D$ denotes a small monetary unit (an empirical parameter), and ${{f}_{i}}$ stands for the probability density of wealth between $\left( i-1 \right) D$ and $i D$.

Next, let us introduce the symbol ${{P}_{i}}$ standing for the probability of wealth between $\left( i-1 \right) D$ and $i D$. Mathematically,
\begin{equation}
	{{P}_{i}} = {{f}_{i}} D
	\label{fi_and_Pi}
	\tt{,}
\end{equation}
and
\begin{equation}
	\sum_{i}{{{P}_{i}}}=1
	\tt{.}
	\label{Sum_Pi}
\end{equation}
From Eqs. (\ref{Entropy_Sum_fi}) - (\ref{Sum_Pi}), we obtain
\begin{equation}
	S=-\sum\limits_{{{P}_{i}}>0}{{{P}_{i}}\ln \left( \frac{{{P}_{i}}}{D} \right)}=-\sum\limits_{{{P}_{i}}>0}{{{P}_{i}}\ln {{P}_{i}}}+\ln D
	\tt{.}
	\label{Entropy_Sum_Pi}
\end{equation}

Remark 1: It is straightforward to observe that Eq. (\ref {Entropy_Integral_f}) aligns with Eqs. (\ref{Entropy_Sum_fi}) and (\ref{Entropy_Sum_Pi}) in the limiting case where $N \to \infty$ and $D \to 0$. However, in real-world societies or virtual systems, these idealized limits are practically unattainable.

Remark 2: The numerical accuracy of entropy calculations can be influenced significantly by the chosen computational method. The error becomes substantial if the number of agents $N$ is relatively small or if the monetary interval $D$ is excessively wide. On the other hand, higher accuracy can be achieved by selecting a sufficiently large $N$ and appropriately narrowing $D$.

Remark 3: The monetary interval $D$ is an empirical parameter that must be carefully calibrated, since it should neither be too large nor too small. As previously noted, the numerical error in entropy calculations using Eq. (\ref{Entropy_Sum_fi}) or Eq. (\ref{Entropy_Sum_Pi}) increases as $D$ becomes larger. Conversely, setting $D$ too small may lead to numerical oscillations. In practice, $D$ is typically determined empirically; in this study, we adopt $D = {m_0} / 100$.

Remark 4: For a uniform economic system where all agents possess the same amount of money, the entropy reaches its minimum value. In this case, the wealth distribution reduces to a Dirac delta distribution:
\begin{equation}
	f(m) = \delta ({m_0})
	\tt{,}
\end{equation}
which satisfies the formula in Eq. (\ref{S1}). For an economic market where each agent owns the same wealth ${m_0}$, the probability distribution of wealth is given by:
\begin{equation}
	{{P}_{i}}=
	\left\{
	\begin{array}{l}
		1, \ \left( i-1 \right) D\le {{m}_{0}}<iD \tt{,}  \\
		0, \ \rm{otherwise} \tt{.}
	\end{array}
	\right.
\end{equation}
As a result, Eq. (\ref{Entropy_Sum_Pi}) simplifies to
\begin{equation}
	S=\ln D
	\tt{,}
	\label{Entropy_Uniform}
\end{equation}
indicating that entropy is solely a function of $D$.

Remark 5: The maximum entropy corresponds to the Boltzmann--Gibbs distribution which characterizes wealth distribution in an idealized free economic system. When external or artificial factors are introduced into a complex market, the wealth distribution deviates from the Boltzmann--Gibbs function. This departure is quantified by a deviation degree greater than zero, resulting in a reduction in entropy. Generally, the entropy decreases as the deviation degree increases.

\section{Numerical validation}\label{SecIII}

There are three critical and intriguing questions that need to be addressed:
(i) Does the total amount of money remain constant across the three models?
(ii) How can the statistical noises in simulations be reduced?
(iii) Can a quasi-steady state be achieved after a sufficiently long period of evolution in these economic markets? To investigate these questions, we consider an economic market with $N=2^{14}=16384$ agents. At the initial time, each agent possesses an equal amount of money, $m_0 = 1$. These agents are further divided into ${N_g} = 128$ distinct groups. Two cases are examined, corresponding to different saving propensities: $\lambda = 0$ and $\lambda = 0.7$.

\begin{figure}[htbp]
	\centering
	\includegraphics[width=0.8\textwidth]{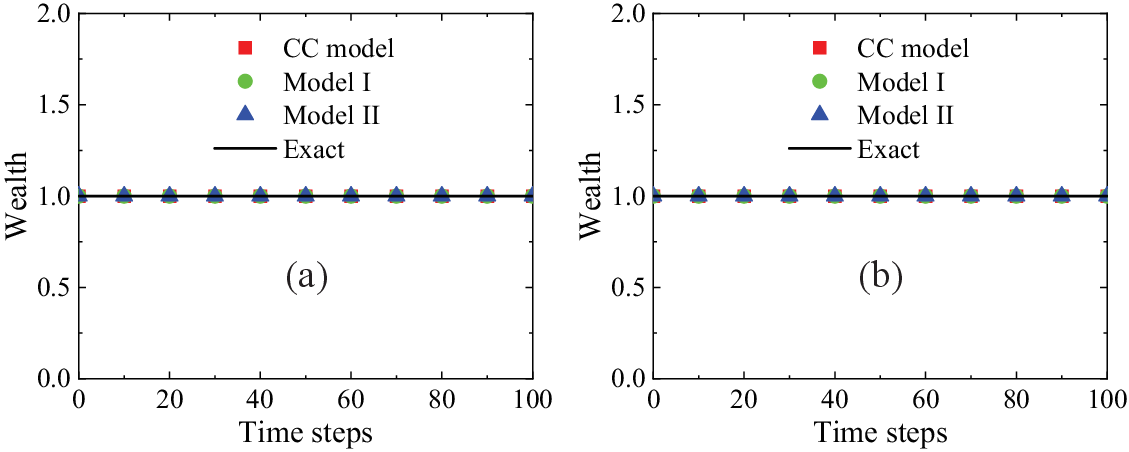}
	\caption{Evolution of average wealth for different saving propensities: (a) $\lambda = 0$ and (b) $\lambda = 0.7$.}
	\label{Fig05}
\end{figure}

From a theoretical standpoint, the principle of money conservation should hold for all three models in a closed market during transactions. To verify this, we compare the simulation results of the three kinetic models with their theoretical predictions. Figure \ref{Fig05} illustrates the evolution of the average wealth over time for the two cases: $\lambda = 0$ and $\lambda = 0.7$. The simulation results of CC model, Model I, and Model II are represented by squares, circles, and triangles, respectively. The solid line corresponds to the exact theoretical solution. As shown in Fig. \ref{Fig05}, the simulation results for all three models perfectly align with the theoretical value throughout the evolution process. This consistency confirms that the total money is conserved across all three models, validating their adherence to the principle of money conservation.

\begin{figure}[htbp]
	\centering
	\includegraphics[width=0.99\textwidth]{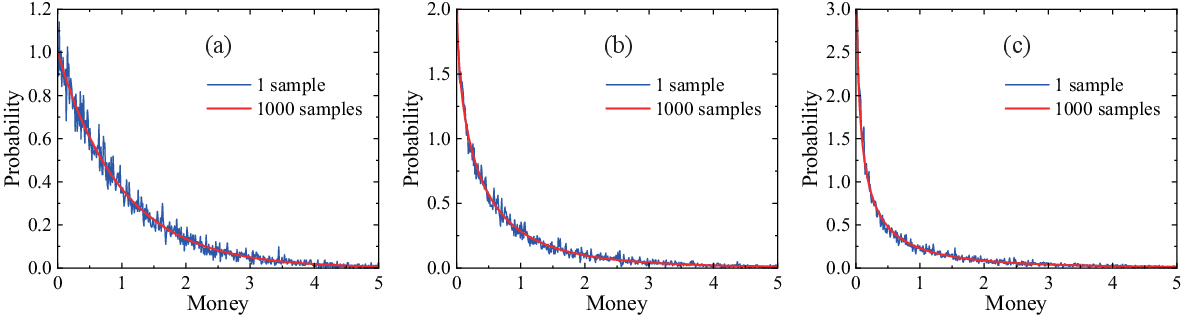}
	\caption{Wealth distributions for a single sample and the average over $1000$ samples: (a) CC model, (b) Model I, and (c) Model II.}
	\label{Fig06}
\end{figure}

Next, we evaluate the impact of statistical noise. In systems with a limited number of agents, statistical fluctuations can obscure the analysis and interpretation of simulation results. To mitigate this issue, one effective approach is to perform multiple simulation runs and calculate the average across all outcomes. To demonstrate this, we compare the results from a single simulation with the averaged results of $1000$ samples. For brevity, only the case of $\lambda = 0$ is considered. Figures \ref{Fig06} (a)-(c) display the wealth distributions for the CC model, Model I, and Model II, respectively. In these figures, the blue line represents the results of a single sample, while the red line depicts the averaged results from $1000$ simulations. It is evident that the wealth distribution from a single sample exhibits significant fluctuations, which can hinder accurate analysis. In contrast, the average over $1000$ samples yields a smooth and reliable curve, effectively reducing noise. Therefore, in the subsequent analyses, we use the averaged results from $1000$ simulations to ensure a more accurate representation of the wealth distribution.

\begin{figure}[htbp]
	\centering
	\includegraphics[width=0.99\textwidth]{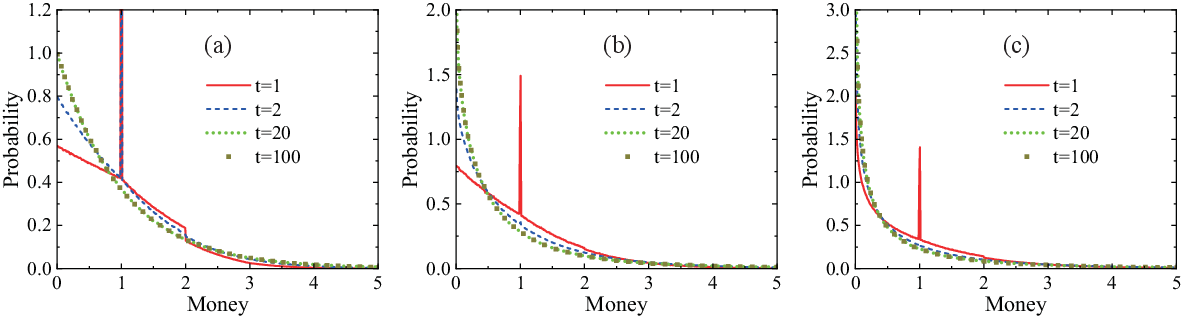}
	\caption{Wealth distributions at various time instants: (a) CC model, (b) Model I, and (c) Model II.}
	\label{Fig07}
\end{figure}

A natural question is whether the wealth distribution reaches a quasi-steady state or continuously changes over time. To address this, we simulate the wealth distribution in an economic market under the condition of ${\lambda} = 0$. Figures \ref{Fig07}(a)-(c) depict the wealth distributions at time steps $t = 1$, $2$, $20$, and $100$ obtained from the CC model, Model I, and Model II, respectively. Initially, the wealth distribution starts as a Dirac delta function. At $t = 1$, the distribution broadens and lowers, reflecting the initial stages of wealth redistribution. By $t = 20$, the wealth distribution converges toward a quasi-steady state, as evidenced by the similarity in profiles at $t = 20$ and $t = 100$. This indicates that a quasi-steady state is effectively reached at $t = 20$. The results demonstrate that the wealth distribution evolves during the initial phase and eventually stabilizes into a steady state after a sufficiently long period. Notably, similar behavior is observed for other values of ${\lambda}$, although these results are omitted here for brevity. In summary, the three kinetic models consistently produce stationary wealth distributions across different saving propensities and collective transaction scenarios.

In the following section, we will explore the characteristics of equilibrium wealth distributions under varying saving propensities and collective transaction conditions.

\section{Numerical investigation}\label{SecIV}

How do saving propensity and collective transactions influence the economic market? Despite their practical significance, the quantitative impact of collective transactions on individuals with saving tendencies has remained unexplored due to the lack of appropriate analytical tools. In this section, we address these questions by utilizing Models I and II to simulate and analyze the economic market under various conditions. The market is modeled with $N = 2^{14} = 16384$ agents, each initially allocated $m_0 = 1$ unit of money. All agents share the same saving propensity $\lambda$ and are evenly divided into $N_g$ distinct groups.

\subsection{Individual and collective transactions: Model I}

As previously mentioned, Model I is designed to capture an economic market that includes both individual transactions among agents with saving tendencies and collective transactions between different groups. Using this model, we simulate the market dynamics and investigate the stationary wealth distribution under varying saving propensities and group configurations.

\begin{figure}[htbp]
	\centering
	\includegraphics[width=0.9\textwidth]{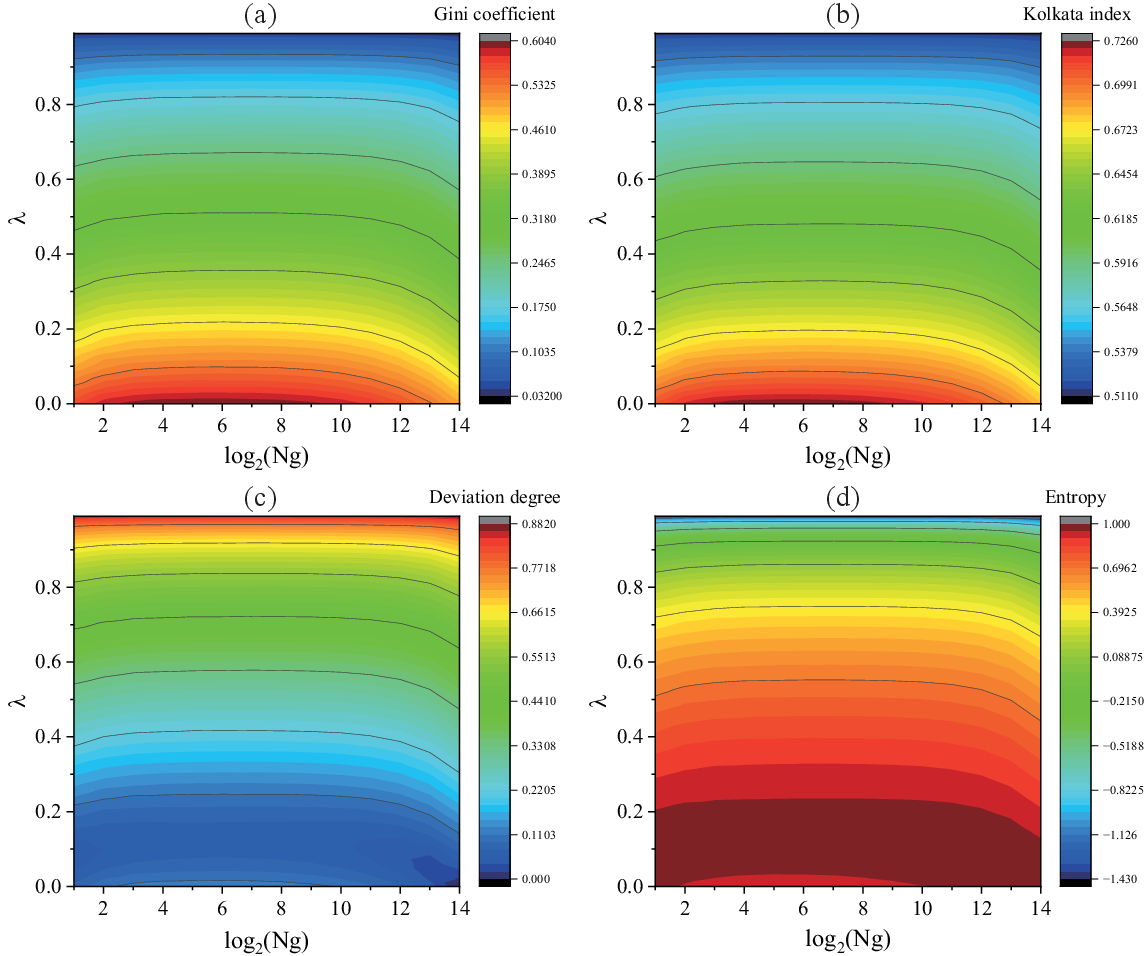}
	\caption{Contours of metrics from Model I in the ($\lambda$, $N_g$) space: (a) Gini coefficient, (b) Kolkata index, (c) deviation degree, and (d) entropy.}
	\label{Fig08}
\end{figure}

To quantitatively analyze the impact of various factors on wealth distributions, Fig. \ref{Fig08} presents contours of metrics from Model I in the space of ($\lambda$, $N_g$), where $0 \le \lambda < 1$ and ${2^0} \le N_g \le {2^{14}}$. Figures \ref{Fig08} (a)-(d) illustrate the Gini coefficient $\rm{G}$, Kolkata index $k$, deviation degree $\Delta$, and entropy $S$, respectively. 

It is observed that the Gini coefficient and Kolkata index exhibit similar patterns of change, while the deviation degree and entropy demonstrate opposite trends. With the increasing of $\lambda$, the Gini coefficient and Kolkata index diminish, whereas the deviation degree (entropy) firstly reduces (rises) and then rises (reduces). Relatively large savings reduce wealth inequality and curb wealth polarization. For example, when there is no saving (${\lambda_i} = 0$), a relatively larger amount of money is exchanged among traders, leading to greater wealth disparities between agents. Conversely, at the maximum saving propensity (${\lambda_i} = 1$), no transactions occur, and all agents retain identical wealth. In summary, saving propensity reduces wealth inequality, regardless of the presence of collective transactions. 

Moreover, Figs. \ref{Fig08} (a) and (b) illustrate that as $N_g$ increases, both the Gini coefficient and the Kolkata index rise until $N_g = {2^7}$ and then begin to decline. In Fig. \ref{Fig08} (c), the deviation degree exhibits a distinct trend: for a small saving propensity, it first increases and then decreases, while for a large saving propensity, the pattern reverses, decreasing initially and then increasing. Similarly, Fig. \ref{Fig08} (d) shows that entropy follows a comparable bifurcated behavior: for a small saving propensity, it decreases at first and then increases, whereas for a large saving propensity, it increases initially before decreasing. 

Notably, Figs. \ref{Fig08} (a)-(d) highlight that all metrics attain their extreme values at $N_g = {2^7}$. These results suggest that an intermediate number of trading groups intensifies wealth inequality, whereas either a very small or an excessively large number of groups fosters greater wealth equality. The critical threshold for the number of groups can be identified as ${{N}_{c}}=\sqrt{N}$, where $N$ represents the total number of agents.

\subsection{Individual and collective transactions: Model II}

As noted earlier, Model II is proposed to simulate an economic market where individual transactions occur among agents within the same group, while collective transactions take place between different groups. In this part, Model II is employed to simulate market dynamics and examine the stationary wealth distribution across various saving propensities and group configurations.

\begin{figure}[htbp]
	\centering
	\includegraphics[width=0.9\textwidth]{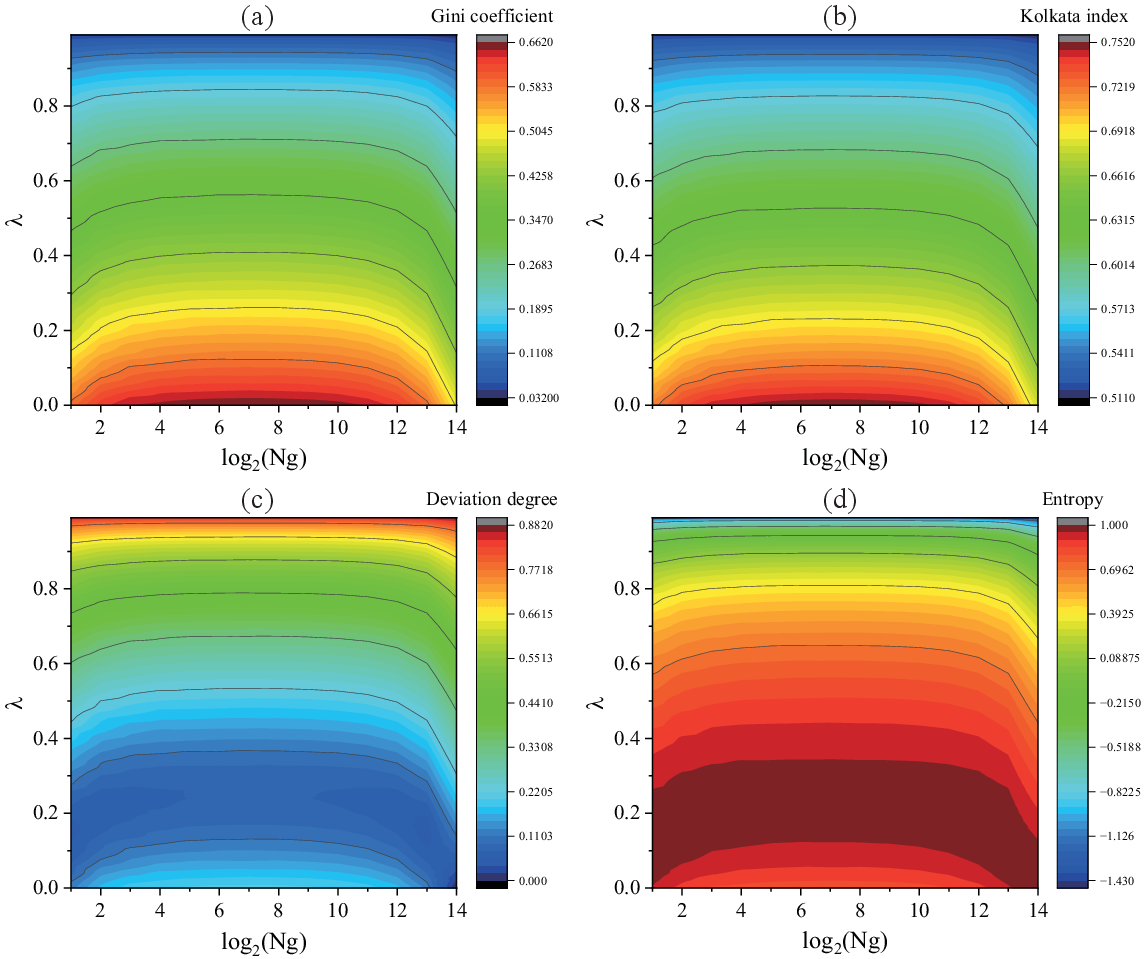}
	\caption{Contours of metrics from Model II in the ($\lambda$, $N_g$) space: (a) Gini coefficient, (b) Kolkata index, (c) deviation degree, and (d) entropy.}
	\label{Fig09}
\end{figure}

Figures \ref{Fig09} (a)-(d) illustrate the contours of the Gini coefficient, Kolkata index, deviation degree, and entropy in the ($\lambda$, $N_g$) parameter space. A comparison between Figs. \ref{Fig08} and \ref{Fig09} reveals that the simulated metrics from Models I and II exhibit similar changing patterns. This similarity suggests that collective transactions play a comparable role in the economic markets simulated by these two kinetic models.

\begin{figure}[htbp]
	\centering
	\includegraphics[width=0.8\textwidth]{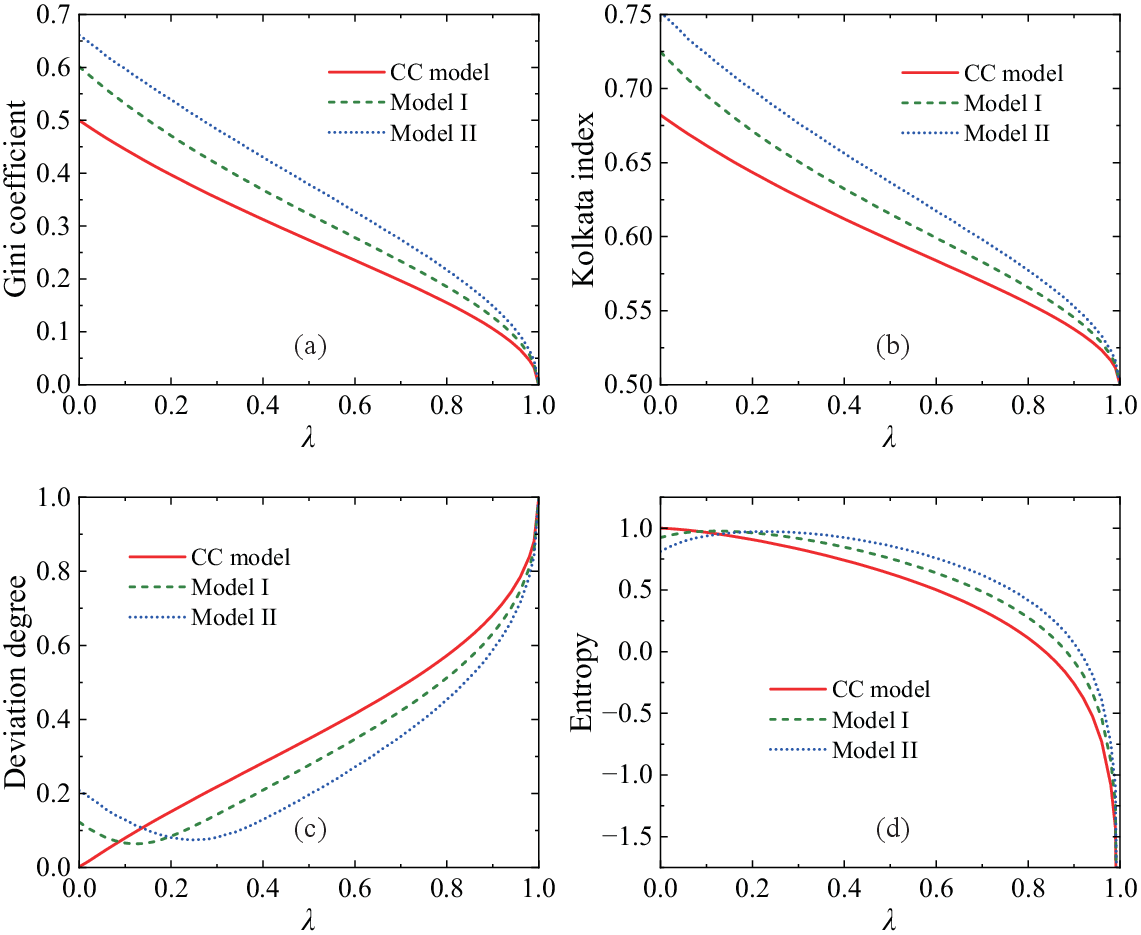}
	\caption{Metrics versus saving propensities: (a) Gini coefficient, (b) Kolkata index, (c) deviation degree, and (d) entropy.}
	\label{Fig10}
\end{figure}

To further investigate, we quantitatively compare the metrics obtained from the CC model, Model I, and Model II, focusing on the case where ${N_g} = 128$. Figures \ref{Fig10} (a)-(d) present the Gini coefficient, Kolkata index, deviation degree, and entropy as functions of the saving propensity.

Figures \ref{Fig10} (a) and (b) show that both the Gini coefficient and the Kolkata index decrease as the saving propensity increases. The Gini coefficients have the same minimum value of $0$, but their maximum values differ: $0.50$, $0.60$, and $0.66$ for the CC model, Model I, and Model II, respectively. Similarly, the Kolkata index reaches a consistent minimum of $0.50$, but its maximum values are $0.68$, $0.72$, and $0.75$ for the CC model, Model I, and Model II, respectively. In both figures, the curve for Model II lies above that of Model I, which in turn is higher than the curve for the CC model. This indicates that collective transactions amplify wealth inequality, whereas individual transactions between agents from different groups help reduce it.

As shown in Fig. \ref{Fig10} (c), for the CC model, the deviation degree increases monotonically from $0$ to $1$, aligning with theoretical predictions. In contrast, for models incorporating collective transactions, the deviation degree initially decreases, reaches a trough, and then increases with saving propensity. These observations highlight three key points:
(i) For relatively low saving propensities, the deviation degree of Model II is higher than that of Model I, which is higher than that of the CC model.
(ii) The saving propensity corresponding to the trough in Model II is larger than that in Model I.
(iii) For relatively high saving propensities, the deviation degree of Model II becomes lower than that of Model I, which is, in turn, lower than that of the CC model.
At the special case of $\lambda = 1$, the deviation degree is zero for all models. These findings can be attributed to the fact that collective transactions exert a more significant influence in Model II than in Model I, as the latter does not allow individual transactions between agents from different groups. The deviation degree is determined by the interplay between saving propensities and the effects of collective transactions.

Finally, in Fig. \ref{Fig10} (d), for the CC model, entropy decreases from $1$ to $-4.6$ as saving interest increases, consistent with theoretical predictions. In models with collective transactions, entropy initially increases, reaches a peak, and then decreases with the saving propensity. A comparison of Figs. \ref{Fig10} (c) and (d) reveals that entropy trends inversely to the deviation degree. Theoretically, as the wealth distribution deviates from the Boltzmann--Gibbs form \cite{Dragulescu2001EPJB}, the deviation degree rises, and entropy declines. When the wealth distribution aligns with the Boltzmann--Gibbs form, the deviation degree is $\Delta = 0$, and entropy reaches its maximum value of $S=1$.

\section{Conclusion}\label{SecV}

In this study, we proposed two kinetic exchange models to simulate the dynamics of a closed economic market with random exchanges, incorporating both saving propensity and collective transactions. Model I captures an economic system that features individual transactions among agents with saving tendencies as well as collective transactions between different groups. Model II simulates a similar market but restricts individual transactions to agents within the same group while allowing collective transactions between groups. To simulate the dual-layered dynamics of these markets, we designed a three-step trading process: (i) collective transactions among groups, (ii) redistribution within groups, and (iii) individual transactions among agents. Saving interest was integrated into the modelling framework using the CC model \cite{Chakraborti2000EPJB}, applied to both individual and collective transactions.

To evaluate the equilibrium states of the economic market, we employed four key metrics: the Gini coefficient, Kolkata index, deviation degree, and entropy. Our results provide new insights into how individual and collective transaction mechanisms interact to shape wealth distribution. Specifically, the simulated Gini coefficients and Kolkata indices indicate that collective transactions tend to amplify wealth inequality by concentrating wealth among groups, whereas individual transactions between agents from different groups mitigate this inequality by redistributing wealth more evenly.

Furthermore, the interplay between saving propensities and collective transactions is found to govern the deviation degree and entropy of the system, which exhibit inverse trends. As the saving propensity increases, the wealth distribution shifts away from the Boltzmann--Gibbs equilibrium, leading to variations of deviation degrees and entropy. Conversely, when the saving propensity aligns with specific thresholds, collective transactions dominate, resulting in unique points characterized by distinct peaks or troughs in these metrics.

Our findings highlight the critical role of dual-layered market interactions in shaping wealth distribution and market dynamics. The proposed models not only enhance our understanding of how collective and individual transactions influence economic inequality but also offer a flexible framework for exploring real-world market phenomena, such as the emergence of wealth concentration and redistribution patterns. Future research can expand on this work by integrating additional factors, such as taxation policies, group heterogeneity, or external shocks, to better reflect the complexities of real economic systems.

\section*{Acknowledgements}

This work is supported by Humanities and Social Science Foundation of the Ministry of Education in China (under Grant No. 24YJCZH163), Guangdong Basic and Applied Basic Research Foundation (under Grant No. 2024A1515010927), National Natural Science Foundation of China (under Grant No. 51806116), and Fundamental Research Funds for the Central Universities, Sun Yat-sen University (under Grant No. 24qnpy044).

\bibliographystyle{tfq}
\bibliography{Reference}

\begin{thebibliography}{10}
\newcommand{\printfirst}[2]{#1}
\newcommand{\switchargs}[2]{#2#1}
\providecommand{\url}[1]{\normalfont{#1}}
\providecommand{\urlprefix}{Available at }

\bibitem{Joohyun2017JS}
J. Kim and D.H. Lee, \emph{{Relative wealth concerns, positive feedback, and
  financial fluctuation}}, Journal of Simulation 11 (2017), pp. 128--136.

\bibitem{Mantegna1999}
R.N. Mantegna and H.E. Stanley, \emph{{Introduction to econophysics:
  correlations and complexity in finance}}, Cambridge university press,
  Cambridge, 1999.

\bibitem{Yakovenko2009RMP}
V.M. Yakovenko and J.B. Rosser Jr., \emph{{Colloquium: Statistical mechanics of
  money, wealth, and income}}, Reviews of Modern Physics 81 (2009), pp.
  1703--1725.

\bibitem{Pareto1897Book}
V. Pareto, \emph{Cours d' economie politique}, Rouge, Lausanne and Paris, 1897.

\bibitem{Gini1912JEI}
C. Gini, \emph{{The origins of the Gini index: extracts from Variabilità e
  Mutabilità (1912) by Corrado Gini}}, Journal of Economic Inequality 10
  (2012), pp. 421--443.

\bibitem{Kuznets1955AER}
S. Kuznets, \emph{Economic growth and income inequality}, American Economic
  Review 45 (1955), pp. 1--28.

\bibitem{Dragulescu2001EPJB}
A. Dr\u{a}gulescu and V.M. Yakovenko, \emph{{Evidence for the exponential
  distribution of income in the USA}}, European Physical Journal B 20 (2001),
  pp. 585--589.

\bibitem{Ghosh2014PA}
A. Ghosh, N. Chattopadhyay, and B.K. Chakrabarti, \emph{{Inequality in
  societies, academic institutions and science journals: Gini and k-indices}},
  Physica A 410 (2014), pp. 30 -- 34.

\bibitem{Chakraborti2000EPJB}
A. Chakraborti and B.K. Chakrabarti, \emph{{Statistical mechanics of money: how
  saving propensity affects its distribution}}, European Physical Journal B 17
  (2000), pp. 167--170.

\bibitem{Chakrabarti2009PA}
A.S. Chakrabarti and B.K. Chakrabarti, \emph{Microeconomics of the ideal gas
  like market models}, Physica A 388 (2009), pp. 4151--4158.

\bibitem{Cha2011CPC}
M.Y. Cha, J.W. Lee, D.S. Lee, and D.H. Kim, \emph{Wealth dynamics in world
  trade}, Computer Physics Communications 182 (2011), pp. 216--218.

\bibitem{Fernandes2020EPJB}
L. Fernandes and J. Tempere, \emph{Effect of segregation on inequality in
  kinetic models of wealth exchange}, European Physical Journal B 93 (2020),
  p.~37.

\bibitem{Munoz2022Chaos}
V. Mu{\~n}oz, \emph{Wealth distribution for agents with spending propensity,
  interacting over a network}, Chaos 32 (2022), p. 123144.

\bibitem{Chatterjee2004PA}
A. Chatterjee, B.K. Chakrabarti, and S. Manna, \emph{{Pareto law in a kinetic
  model of market with random saving propensity}}, Physica A 335 (2004), pp.
  155--163.

\bibitem{Cerda2013MCM}
J. Cerd{\'a}, C. Montoliu, and R. Colom, \emph{Lgem: A lattice boltzmann
  economic model for income distribution and tax regulation}, Mathematical and
  Computer Modelling 57 (2013), pp. 1648--1655.

\bibitem{Angle1986SF}
J. Angle, \emph{{The Surplus Theory of Social Stratification and the Size
  Distribution of Personal Wealth}}, Social Forces 65 (1986), pp. 293--326.

\bibitem{Boghosian2017PA}
B.M. Boghosian, A. Devitt-Lee, M. Johnson, J. Li, J.A. Marcq, and H. Wang,
  \emph{Oligarchy as a phase transition: The effect of wealth-attained
  advantage in a fokker--planck description of asset exchange}, Physica A 476
  (2017), pp. 15--37.

\bibitem{Vermeulen2018RIW}
P. Vermeulen, \emph{How fat is the top tail of the wealth distribution?},
  Review of Income and Wealth 64 (2018), pp. 357--387.

\bibitem{During2018TEPJB}
B. D{\"u}ring, L. Pareschi, and G. Toscani, \emph{Kinetic models for optimal
  control of wealth inequalities}, European Physical Journal B 91 (2018), p.
  265.

\bibitem{Li2019PA}
J. Li, B.M. Boghosian, and C. Li, \emph{The affine wealth model: An agent-based
  model of asset exchange that allows for negative-wealth agents and its
  empirical validation}, Physica A 516 (2019), pp. 423--442.

\bibitem{Cui2020Entropy}
L. Cui and C. Lin, \emph{Lattice--gas--automaton modeling of income
  distribution}, Entropy 22 (2020), p. 778.

\bibitem{Cui2021PA}
L. Cui and C. Lin, \emph{{A simple and efficient kinetic model for wealth
  distribution with saving propensity effect: Based on lattice gas automaton}},
  Physica A 561 (2021), p. 125283.

\bibitem{Sargent2021PNAS}
T.J. Sargent, N. Wang, and J. Yang, \emph{Earnings growth and the wealth
  distribution}, Proceedings of the National Academy of Sciences of the United
  States of America 118 (2021), p. e2025368118.

\bibitem{Goswami2022PTRSA}
S. Goswami, \emph{A poor agent and subsidy: an investigation through ccm
  model}, Philosophical Transactions of the Royal Society A 380 (2022), p.
  20210166.

\bibitem{Patriarca2004PA}
M. Patriarca, A. Chakraborti, and K. Kaski, \emph{{Gibbs versus non-Gibbs
  distributions in money dynamics}}, Physica A 340 (2004), pp. 334--339.

\bibitem{Patriarca2007EPJB}
M. Patriarca, A. Chakraborti, E. Heinsalu, and G. Germano, \emph{{Relaxation in
  statistical many-agent economy models}}, European Physical Journal B 57
  (2007), pp. 219--224.

\bibitem{Chakraborti2008PJOP}
A. Chakraborti and M. Patriarca, \emph{{Gamma-distribution and wealth
  inequality}}, Pramana - Journal of Physics 71 (2008), pp. 233--243.

\bibitem{Patriarca2010EPJB}
M. Patriarca, E. Heinsalu, and A. Chakraborti, \emph{{Basic kinetic
  wealth-exchange models: common features and open problems}}, European
  Physical Journal B 73 (2010), pp. 145--153.

\bibitem{Chakraborti2011QF}
A. Chakraborti, I.M. Toke, M. Patriarca, and F. Abergel, \emph{{Econophysics
  review: II. Agent-based models}}, Quantitative Finance 11 (2011), pp.
  1013--1041.

\bibitem{Heinsalu2014EPJB}
E. Heinsalu and M. Patriarca, \emph{{Kinetic models of immediate exchange}},
  European Physical Journal B 87 (2014).

\bibitem{Chatterjee2017PA}
A. Chatterjee, A. Ghosh, and B.K. Chakrabarti, \emph{{Socio-economic
  inequality: Relationship between Gini and Kolkata indices}}, Physica A 466
  (2017), pp. 583--595.

\bibitem{Patriarca2017book1}
M. Patriarca, E. Heinsalu, A. Singh, and A. Chakraborti, \emph{{Kinetic
  Exchange Models as D Dimensional Systems: A Comparison of Different
  Approaches}}, 2017.

\bibitem{Patriarca2017book2}
M. Patriarca, E. Heinsalu, A. Chakraborti, and K. Kaski, \emph{{The Microscopic
  Origin of the Pareto Law and Other Power-Law Distributions}}, 2017.

\bibitem{Hu2006EPJB}
M.B. Hu, W.X. Wang, R. Jiang, Q.S. Wu, B.H. Wang, and Y.H. Wu, \emph{A unified
  framework for the pareto law and matthew effect using scale-free networks},
  European Physical Journal B 53 (2006), pp. 273--277.

\bibitem{Ghosh2016PA}
A. Ghosh, A. Chatterjee, J.i. Inoue, and B.K. Chakrabarti, \emph{Inequality
  measures in kinetic exchange models of wealth distributions}, Physica A 451
  (2016), pp. 465--474.

\bibitem{Kulp2019Chaos}
C.W. Kulp, M. Kurtz, N. Wilston, and L. Quigley, \emph{{The effect of various
  tax and redistribution models on the Gini coefficient of simple exchange
  games}}, Chaos 29 (2019), p. 083118.

\bibitem{Cui2023IJMPC}
L. Cui and C. Lin, \emph{Kinetic modeling of economic markets with
  heterogeneous saving propensities}, International Journal of Modern Physics C
  34 (2023), p. 2350106.

\bibitem{Cui2023EPL}
L. Cui, C. Lin, and X. Huang, \emph{{Kinetic modeling of wealth distribution
  with saving propensity, earnings growth and Matthew effect}}, EPL 143 (2023),
  p. 12002.

\bibitem{Banerjee2020FIP}
S. Banerjee, B.K. Chakrabarti, M. Mitra, and S. Mutuswami, \emph{{Inequality
  Measures: The Kolkata Index in Comparison With Other Measures}}, Frontiers in
  Physics 8 (2020), p. 562182.

\bibitem{Banerjee2023IJMPC}
S. Banerjee, S. Biswas, B.K. Chakrabarti, S.K. Challagundla, A. Ghosh, S.R.
  Guntaka, H. Koganti, A.R. Kondapalli, R. Maiti, M. Mitra, and D.R.S. Ram,
  \emph{{Evolutionary dynamics of social inequality and coincidence of Gini and
  Kolkata indices under unrestricted competition}}, International Journal of
  Modern Physics C 34 (2023), p. 2350048.

\end{thebibliography}


\end{document}